\documentclass{emulateapj}
\usepackage{ulem}
\usepackage{longtable}
\usepackage{graphicx}
\input{epsf.sty}

\shorttitle{Extending the Correlation of $L_R - L_X$ to Gamma Ray Bursts}
\shortauthors{L\"u et al.}

\begin{document}
\title{Extending the Correlation of $L_R - L_X$ to Gamma Ray Bursts}

\author{Jing L\"u\altaffilmark{1,2},  Jing-Wen Xing\altaffilmark{3}, Yuan-Chuan Zou\altaffilmark{3}, Wei-Hua Lei\altaffilmark{3}, Qingwen Wu\altaffilmark{3}, and Ding-Xiong Wang\altaffilmark{3}}

\altaffiltext{1}{GXU-NAOC Center for Astrophysics and Space Sciences, Department of Physics, Guangxi University,  Nanning 530004, China}
\altaffiltext{2}{Guangxi Key Laboratory for Relativistic Astrophysics,  Nanning 530004, China}
\altaffiltext{3}{School of Physics, Huazhong University of Science and Technology, Wuhan, 430074, China. Email: zouyc@hust.edu.cn (YCZ)}

\begin{abstract}
The well-known correlation between the radio luminosity  ($L_R$) and the X-ray luminosity ($L_X$)  $L_R / L_X \simeq 10^{-5}$ holds for a variety of objects like active galactic nuclei, galactic black holes,  solar flares and cool stars. Here we extend the relation to gamma-ray bursts (GRBs), and find the GRBs also lay on the same $L_R-L_X$ relation, with a slightly different slope as $L_R \propto L_X^{1.1}$. This relation implies the explosions in different scales may have common underlying origin.
\end{abstract}


\section{Introduction}
\label{intro}
With the enhancement of the observational instruments, more and more astronomical objects have been discovered continuum radiation from the radio to the X-rays, including active galactic nuclei (AGN) \citep{Sanders1989, Merloni2003}, the Sun, cool star \citep{Guedel1993} and gamma ray burst (GRB) \citep{Kumar2015}, for instance. Recently, the relation between the radio and X-ray emission among different scale objects obtained much attention.
The radio flares and soft X-rays of dwarf M-star UV Ceti were observed simultaneously by VLA and ROSAT/HRI telescope respectively\citep{Benz1996}. For active cool stars, \citet{Guedel1993} gave a tight correlation between the quiescent radio and X-ray emission, where $L_R/L_X \sim 10^{-5}$, known as the G$\ddot{u}$del-Benz relation. During the ROSAT All-Sky Survey accompanied by mostly simultaneous VLA observations, \citet{Guedel1993b} also gave a tight correlation between radio and X-ray luminosities of M dwarfs. \citet{Benz1994} showed the soft X-ray and radio luminosities of solar flares and the coronae of main-sequence, late-type stars follow the relation, $L_{\nu,X}/L_{\nu,R}=k\times10^{15.5\pm0.5}$, where $k$ is a coefficient and different types of stars have different values. \citet{Linsky1996} suggested that the heating mechanism of stellar corona is a flare-like process, and in the review, \citet{van1999} discussed various aspects of stellar flares. \citet{Guedel1996} discussed a few aspects of radio and X-ray studies of solar-like stars, and gave that nonthermal radio and thermal soft X-ray emissions from solar-like stars provided direct information on particle acceleration and coronae heating in the magnetically confined outer atmospheres. \citet{Laor2008} found that there is a more tight correlation between the radio and X-ray fluxes of accretion disks in radio-quiet quasars as in coronally active stars, $L_R/L_X \lesssim 10^{-5}$ (with a 1$\sigma$ scatter of about a factor of 3 in $L_R$). They proposed that most of the radio emission might come from coronae, and the coronae are magnetically heated. \citet{Middleton2013} showed that the X-ray and radio emission are coupled in some sources. Continued radio and X-ray monitoring of some sources should reveal the causal relationship between the accretion flow and the powerful jet emission.  Here we try to extend the $L_R - L_X$ relation  to GRBs.

GRBs are high energy $\gamma$-rays lasting from several milliseconds to a few thousand seconds, and they are the most powerful explosions in the universe \citep{Piran2004, Zhang2007}. GRBs were discovered in the late 1960s by the military Vela satellites \citep{Klebesadel1973}, and caused the great interests of people afterwards. Because the duration of the GRB explosion is very short, the precise location can not be determined in the beginning. Also BeppoSAX satellite, launched on 30 April 1996, confirmed the cosmological distance of GRBs in 1997 \citep{Kumar2015}. With the launching of space missions (such as {\it Swift} and {\it Fermi}), they have greatly enriched our knowledge of this phenomenon. With the more and more observational data, statistical studies have summarized some empirical relationship, such as $E_{peak}\sim E_{\gamma,iso}$ \citep{Amati2002}, $E_{peak}\sim L_{\gamma,iso}$ \citep{Yonetoku2004}, $E_{peak}\sim E_{\gamma}$ \citep{Ghirlanda2004}, $E_{peak}\sim E_{\gamma,iso}\sim t_b$ \citep{Liang2005}, $\Gamma_{0}\sim E_{\gamma,iso}$ \citep{Liang2010} and $\Gamma_{0}\sim L_{\gamma,iso}$ \citep{Ghirlanda2012,Lv2012}.
 A dedicated theoretical study for the $\Gamma_0 - L_{\gamma, iso}$ relation suggests that it can be a justification of the central engine models of GRBs \citep{Lei2013}. Recently, \citet{Wu2011} found a uniform correlation between the synchrotron luminosity and Doppler factor for both GRBs and blazers, which implies that they may share some similar jet physics. The recent work by \citet{Wang2014} on Sw J1644+57 (a tidal disruption event candidate which launches a relativistic jet) may support this similarity. In their work, the radio light curve of Sw J1644+57 were the successfully  interpreted with the afterglow jet model.
The discovery of the long-lived afterglows at X-ray, optical and radio wavelengths advanced the study of large samples of GRBs. \citet{Chandra2012} gave a catalog of radio afterglow of GRBs over a 14 year period from 1997 to 2011, and also include the 11hr X-ray observations in it. These progresses make it possible to extend the  $L_R - L_X$ relation  to GRBs.

This paper is organized as follows. In section \ref{main}, we give the samples selection and analysis of solar flares, cool stars, AGNs and GRBs data. We end with brief conclusion and discussion in section \ref{conclusion}.

\section{Data and analysis}
\label{main}

We investigate the $L_R \sim L_X$ relation of objects in different scales, including stars, AGNs, and GRBs. The data of the solar flares and cool stars are taken from \citet{Laor2008}, who tabulated the data of \citet{Benz1994}. Solar flare soft X-rays are conventionally measured in a narrow and harder band than the stellar sources, and the data points from the solar flares are for different classes that may deviate from the overall linear relation \citet{Benz1994}.  We take the $L_R$ and $L_X$ of AGNs from \citet{Merloni2003}.

Our analysis focus on $L_R$ and $L_X$ for GRBs. Based on the radio peak flux densities and luminosity distances $D_L$ calculated from the redshift $z$, we can obtain the radio luminosity of GRBs afterglow, $L_R \simeq 4 \pi D_L^{2} F_{\nu, m} \times \nu_R $.
For GRBs, we chose 47 samples with both radio and X-ray observations. They are listed in Table 1. Before GRB 111215A, 42 GRBs are taken from \citet{Chandra2012}.  Also we can obtain bursts radio frequency and peak flux densities in their Table 3, X-ray flux at 11hr after burst in their Table 6. For the other five bursts we have used: GRB 111215A: \citet{van der Horst2015} for redshift, \citet{Zauderer2013} for radio peak flux density, and in it's Figure 3, there are five radio light curves at 5.8, 8.4, 19.1, 24.4 and 93GHz, but from their profiles, the first three may result from the positive and negative shock, and the remaining two may result from external shock, so we chose the peak flux density of 8.4GHz. GRB 120326A: \citet{Kruehler2012} for redshift, \citet{Staley2013} for radio peak flux density. GRB 130427A: \citet{Xu2013} and \citet{Flores2013} for redshift, \citet{Anderson2014} for radio peak flux density. GRB 130603B: \citet{Cucchiara2013} for redshift, \citet{Fong2014} for radio peak flux density. GRB 130907A: \citet{de Ugarte Postigo2013} for redshift, \citet{Veres2014} for radio peak flux density. The X-ray flux at 11hr of these five bursts are taken from Swift-XRT products for GRBs \footnote{http://www.swift.ac.uk/xrt\_products/index.php}.

In figure 1, we give the distribution of the 15 solar flares (black dot circles), 65 cool stars (blue stars), 149 AGNs (red triangles) and 47 GRBs (black solid circles) objects in the $L_R \sim L_X$ plane. The black solid line in lower panel, $L_R \cong 3.42 \times 10^{-10} L_X^{1.122}$ is the best fit, and compared to the red dashed line, $L_R = 10^{-5}L_X$. Remarkably, all of them appear to follow a similar $L_R \sim L_X$ relation, despite they span $\sim 30$ orders of magnitude in luminosity between these different types of active objects.


We investigate the $L_R \sim L_X$ relation mentioned by \citet{Benz1994} with our larger samples. \citet{Laor2008} gave the linear relation $\log{L_{R,39}}=(-0.21\pm0.08)+(1.08\pm0.15)\log{L_{X,44}}$ for radio quiet quasar (RQQ). In Figure 1, we plot the $L_R$ and $L_X$ on $\log~\log$ coordinates of our 276 samples, and gets a slightly modified relation
\begin{equation}
\label{eq:total}
   \log{L_R}=(-9.466\pm0.102)+(1.122\pm0.002)\log{L_X}
\end{equation}
with a correlation coefficient $\zeta=0.980$. As a comparison, we have also selected all the 47 GRBs, trying to examine the correlation between $L_R$ and $L_X$, the best fitting result is
\begin{equation}
   \log{L_R}=(9.288\pm1.339)+(0.695\pm0.028)\log{L_X}
\end{equation}
with a correlation coefficient $\zeta=0.814$.

These correlations can be translated to
\begin{equation}
   L_R\simeq 3.42 \times 10^{-10}L_X^{1.122},
\end{equation}
and
\begin{equation}
\label{eq:grb}
   L_R\simeq 1.94 \times 10^{9}L_X^{0.695}.
\end{equation}
The correlation Eq.(\ref{eq:total}) coincide with \citet{Laor2008}, and also can be seen that the $L_X/L_R=k\times10^{15.5\pm0.5}$ correlation discovered by \citet{Benz1994} is confirmed in Eq.(\ref{eq:grb}) with a slightly different slope. The slope of this correlation becomes flatter when the samples are expanded.

\begin{figure}
\begin{center}
\includegraphics[width=0.5\textwidth]{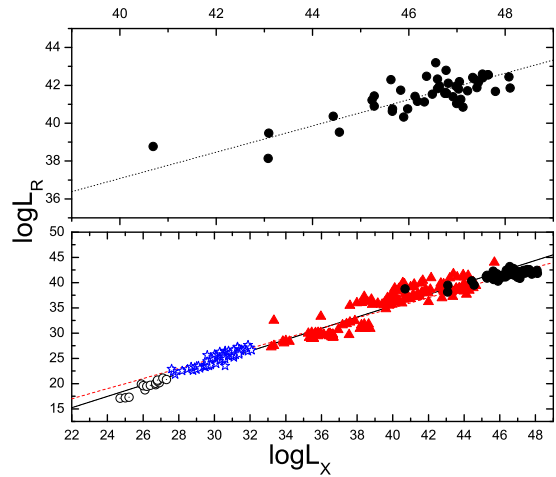}
 \caption{The upper panel includes only the GRBs and their best fitting liner correlation in black dotted line ($L_R \simeq 1.94 \times 10^{9}L_X^{0.695}$). The lower panel contains 15 solar flares (black dot circles), 65 cool stars (blue stars), 149 AGNs (red triangles) and 47 GRBs (black solid circles) objects in the $L_R-L_X$ plane. The black solid line, $L_R \cong 3.42 \times 10^{-10} L_X^{1.122}$ is the best fit. The red dashed line indicates $L_R = 10^{-5}L_X$.}
\end{center}
\label{fig:GE}
\end{figure}

\section{Conclusion and Discussion}
\label{conclusion}
The search of $L_R$ and $L_X$ in GRBs was done in this work. We found the $L_R - L_X$ correlation for solar, cool stars and AGNs can be extended to GRBs.
Even astrophysical phenomenons are divided into different classes, they still have common features on some properties. That implies they essentially may have some commonality. \citet{Franciosini1995} presented a model of magnetic loop which only have one free parameter, the thermal plasma density, indicating a possible co-spatiality of the two emissions, radio emission and X-ray emission. On the other hand, the positive correlation between $L_R$ and $L_X$ is natural.
However, the detailed underlying nature of this relation is still unclear.

Not like the other objects, the radio and X-ray data of GRBs are not simultaneous. The radio peak of GRB afterglows are generally at several days or even months, while the X-ray data are frozen to be at the 11 hours of the observer's frame. One reason is that the simultaneous data is very rare. Thanks to the fast corresponding of {\it Swift} XRT, the X-ray data can be obtained very quickly. But it also decays quickly, and fades out in a few days. There is no fast slewing radio telescope to observe the very early radio emission. Also notice the radio emission is alway optically thick in the early stage, its luminosity is depressed and sensitively related to its size.
Therefore, even occasionally the GRB location is inside the field of view of a radio telescope, it is very hard to be detected because of the faint luminosity. Later on, with the expansion of the GRB ejecta, the radio emission emerges while the X-ray fades away.
That means the simultaneous radio and X-ray luminosities may not obey the proportional relation. From the light curves of GRB X-ray afterglows, the X-ray luminosity at 11 hours generally represents a typical X-ray emission.  The correlation of the peak radio luminosity and X-ray luminosity at 11 hours may indicate their energy source origin should be related, rather than microphysical radiation mechanisms. This extension to the various objects may also indicates the radio and X-ray emission may have the same engine while presenting in two different bands.

The fundamental plane between $L_X$, $L_R$ and $M_{BH}$ is even tighter in stellar mass and supermassive black holes \citep{Merloni2003}. Though the GRB is believed harboring a stellar mass black hole, the fundamental plane is not extending to GRBs.

\acknowledgments
 This work is supported by the National Natural Science Foundation of China (Grants No. U1231101 and 11173011),  and the Guangxi Science Foundation (Grant No. 2014GXNSFBA118009).


%

\begin{deluxetable}{lrrrrrrrr}
\tablewidth{0pt}
\tablecaption{The redshift, radio luminosity and X-ray luminosity of the GRBs. \label{Tab:publ-works}}
\tablehead{\colhead{GRB}  &    \colhead{$z$}  &  \colhead{$D_L^{\mathrm a}$}  &  \colhead{$\nu_R^{\mathrm b}$}  &  \colhead{$F_m^{\mathrm c}$}  &  \colhead{$F_R^{\mathrm d}$}  &  \colhead{$F^{11h\,\mathrm e}_X$} & \colhead{$L_R^{\mathrm f}$}  &  \colhead{$L_X^{\mathrm g}$} }
\startdata
970508  & 0.835 & 5299.32  & 8.46   & 958   & 8.10     & 5.7   & 27.23   & 1.92     \\
970828  & 0.958 & 6288.80  & 8.46   & 144   & 1.22     & 19.9  & 5.76    & 9.42     \\
980425  & 0.009 & 38.27    & 8.64   & 39360 & 340.07   & 2.8   & 0.060   & 0.000049 \\
980703  & 0.966 & 6354.40  & 8.46   & 1370  & 11.59    & 14.0    & 56.0    & 6.76     \\
981226  & 1.110 & 7558.88  & 8.46   & 137   & 1.16     & 2.8   & 7.92    & 1.91     \\
990510  & 1.619 & 12108.32 & 8.46   & 255   & 2.16     & 34.7  & 37.84   & 60.87    \\
991216  & 1.020 & 6800.95  & 15.00  & 1100  & 16.50    & 78.1  & 91.31   & 43.22    \\
010222  & 1.477 & 10800.10 & 8.46   & 93    & 0.79     & 70.5  & 10.98   & 98.39    \\
011211  & 2.140 & 17105.31 & 8.46   & 162   & 1.37     & 180.0   & 47.98   & 630.15   \\
021004  & 2.330 & 18991.03 & 22.50  & 1614  & 36.32    & 8.4   & 1567.09 & 36.25    \\
030226  & 1.986 & 15599.76 & 22.50  & 328   & 7.38     & 13.5  & 214.88  & 39.31    \\
030329  & 0.169 & 804.57   & 43.00  & 59318 & 2550.67  & 548.6 & 197.56  & 4.25     \\
031203  & 0.105 & 479.23   & 22.50  & 483   & 10.87    & 4.5   & 0.30    & 0.012    \\
050401  & 2.898 & 24785.55 & 8.46   & 122   & 1.03     & 35.1  & 75.86   & 258.00   \\
050416A & 0.650 & 3887.37  & 8.46   & 373   & 3.16     & 25.3  & 5.71    & 4.57     \\
050525A & 0.606 & 3566.96  & 8.46   & 164   & 1.39     & 51.5  & 2.11    & 7.84     \\
050603  & 2.821 & 23987.77 & 8.46   & 377   & 3.19     & 33.0    & 219.59  & 227.20   \\
050730  & 3.968 & 36185.46 & 8.46   & 212   & 1.79     & 76.7  & 280.99  & 1201.64  \\
050820A & 2.615 & 21871.45 & 8.46   & 150   & 1.27     & 221.2 & 72.63   & 1266.05  \\
050824  & 0.830 & 5259.90  & 8.46   & 152   & 1.29     & 13.7  & 4.26    & 4.54     \\
050904  & 6.290 & 62311.66 & 8.46   & 76    & 0.64     & 0.5   & 298.70  & 23.23    \\
051022  & 0.809 & 5095.05  & 8.46   & 268   & 2.27     & 426.5 & 7.04    & 132.47   \\
060218  & 0.033 & 142.97   & 22.5   & 250   & 5.63     & 48.8  & 0.014   & 0.012    \\
060418  & 1.490 & 10918.75 & 8.46   & 216   & 1.83     & 9.4   & 26.07   & 13.41    \\
070125  & 1.548 & 11450.89 & 22.50  & 1778  & 40.01    & 37.8  & 627.63  & 59.30    \\
071003  & 1.604 & 11968.90 & 8.46   & 616   & 5.21     & 55.7  & 89.32   & 95.47    \\
071010B & 0.947 & 6198.94  & 8.46   & 341   & 2.89     & 45.6  & 13.26   & 20.97    \\
071020  & 2.146 & 17164.39 & 8.46   & 141   & 1.19     & 16.1  & 42.05   & 56.75    \\
080603A & 1.687 & 12743.86 & 8.46   & 207   & 1.75     & 15.8  & 34.03   & 30.70    \\
090313  & 3.375 & 29800.55 & 8.46   & 435   & 3.68     & 31.9  & 391.04  & 338.96   \\
090323  & 3.570 & 31883.30 & 8.46   & 243   & 2.06     & 28.1  & 250.04  & 341.78   \\
090328  & 0.736 & 4531.37  & 8.46   & 686   & 5.80     & 61.4  & 14.26   & 15.08    \\
090423  & 8.260 & 85417.81 & 8.46   & 50    & 0.42     & 5.0     & 369.27  & 436.49   \\
090424  & 0.544 & 3126.69  & 8.46   & 236   & 2.00     & 2.3   & 2.34    & 0.27     \\
090715B & 3.000 & 25847.66 & 8.46   & 191   & 1.62     & 8.2   & 129.17  & 65.55    \\
090902B & 1.883 & 14605.36 & 8.46   & 84    & 0.71     & 47.0    & 18.14   & 119.96   \\
091020  & 1.710 & 12960.08 & 8.46   & 399   & 3.38     & 19.5  & 67.84   & 39.19    \\
100414A & 1.368 & 9814.85  & 8.46   & 524   & 4.43     & 143.7 & 51.10   & 165.63   \\
100418A & 0.620 & 3668.21  & 8.46   & 1218  & 10.30    & 10.7  & 16.59   & 1.72     \\
100814A & 1.440 & 10463.72 & 7.90   & 613   & 4.84     & 82.0    & 63.44   & 107.42   \\
100901A & 1.408 & 10174.38 & 33.60  & 378   & 12.70    & 90.1  & 157.31  & 111.60   \\
100906A & 1.727 & 13120.28 & 8.46   & 215   & 1.82     & 27.1  & 37.46   & 55.82    \\
111215A & 2.060 & 16320.50 & 8.40   & 1340  & 11.26    & 140.0 & 358.73  & 446.30   \\
120326A & 1.798 & 13792.90 & 15.00  & 771   & 11.57    & 93.2 & 263.25  & 212.10    \\
130427A & 0.340 & 1783.10  & 15.70  & 4183  & 65.67    & 2167.6  & 24.98  & 82.46    \\
130603B & 0.3565& 1882.80  & 6.70   & 118.6 & 0.79     & 8.5  & 0.34    & 0.36     \\
130907A & 1.238 & 8663.80  & 15.00  & 1060  & 15.90    & 308.1 & 142.80  & 276.74 \\
\enddata
\tablecomments{
a. Luminosity distance in unit of Mpc.
b. The frequency of radio band in unit of  GHz.
c. The radio peak flux densities in  unit of $\mathrm {\mu Jy}$.
d. The radio peak flux in  unit of $10^{-20} \,\mathrm {erg\, cm^{-2}\, s^{-1}}$.
e. X-ray flux at 11hr in  unit of $10^{-13} \,\mathrm {erg\, cm^{-2}\,s^{-1}}$. The fluxes are measured in 0.3-10 ${\rm keV}$ energy range.
f. The radio luminosity of radio band, in  unit of $10^{40}\,\mathrm { erg \, s^{-1}}$.
g. The 11hr X-ray luminosity, in  unit of $10^{45}\, \mathrm {erg \, s^{-1}}$.
}
\end{deluxetable}


\begin{thebibliography}{}
\bibitem[{Amati et al.}(2002)]{Amati2002} Amati, L., Frontera, F., Tavani, M. 2002, A\&A, 390, 81
\bibitem[{Anderson et al.}(2014)]{Anderson2014} Anderson, G. E., van der Horst, A. J., Staley, T. D. et al. 2014 MNRAS, 440, 2059
\bibitem[{Benz \& G$\ddot{u}$del}(1994)]{Benz1994} Benz A., O., G$\ddot{u}$del M. 1994, A\&A, 285, 621
\bibitem[{Benz et al.}(1996)]{Benz1996} Benz A., O., G$\ddot{u}$del M. \& Schmitt, J. H. M. M. 1996, ASPC, 93, 291
\bibitem[{Chandra \& Frail}(2012)]{Chandra2012} Chandra, P. \& Frail, D. A. 2012, \apj, 746, 156
\bibitem[{Cucchiara et al.}(2013)]{Cucchiara2013} Cucchiara, A., Prochaska, J. X., Perley, D. et. al. 2013, \apj, 777, 94
\bibitem[{de Ugarte Postigo et al.}(2013)]{de Ugarte Postigo2013} de Ugarte Postigo, A., Xu, D., Malesani, D., Gorosabel, J., Jakobsson, P., \& Kajava, J. 2013, GRB Coordinates Network, 15187, 1
\bibitem[{Flores et al.}(2013)]{Flores2013} Flores, H. et al., 2013, GCN Circ, 14491
\bibitem[{Fong et al.}(2014)]{Fong2014} Fong, W., Berger, E., Metzger, B. D. et. al. 2014, \apj, 780, 118
\bibitem[{Franciosini \& Chiuderi Drago}(1995)]{Franciosini1995} Franciosini, E. \& Chiuderi Drago, F. 1995, A\&A, 297, 535

\bibitem[{Ghirlanda et al.}(2004)]{Ghirlanda2004} Ghirlanda Giancarlo, Ghisellini Gabriele, Lazzati Davide. 2004, \apj, 616, 331
\bibitem[Ghirlanda et al.(2012)]{Ghirlanda2012}Ghirlanda, G., Nava, L., Ghisellini, G., et al. 2012, MNRAS, 420, 483
\bibitem[{G$\ddot{u}$del \& Benz}(1993)]{Guedel1993} G$\ddot{u}$del, M., Benz, A. O. 1993, \apj, 405, 63
\bibitem[{G$\ddot{u}$del et al.}(1993b)]{Guedel1993b} G$\ddot{u}$del, M., Schmitt, J. H. M. M., Bookbinder, J. A. \& Fleming, T. A. 1993b, \apj, 415, 236
\bibitem[{G$\ddot{u}$del \& Benz}(1996)]{Guedel1996} G$\ddot{u}$del, M. 1996, IAUS, 176, 485
\bibitem[{Klebesadel et al.}(1973)]{Klebesadel1973} Klebesadel, R. W., Strong, I. B., Olson, R. A., Jun. 1973, Observations of Gamma-Ray Bursts of Cosmic Origin, \apj, 182, L85.
\bibitem[{Kruehler et al.}(2012)]{Kruehler2012} Kruehler, T., Fynbo, J. P. U., Milvang-Jensen, B., Tanvir, N., \& Jakobsson, P. 2012, GCN, 13134, 1
\bibitem[{Kumar \& Zhang}(2015)]{Kumar2015} Kumar, Pawan, Zhang, Bing. 2014, arXiv1410.0679K
\bibitem[{Laor \& Behar}(2008)]{Laor2008} Laor, A., Behar, E. 2008, MNRAS, 390, 847
\bibitem[Lei, Zhang \& Liang(2013)]{Lei2013}Lei, W.-H., Zhang B. \& Liang E.-W., ApJ, 765, 125
\bibitem[Liang et al.(2010)]{Liang2010}Liang, E.-W., Yi, S.-X., Zhang, J., et al. 2010, ApJ, 725, 2209
\bibitem[{Liang \& Zhang}(2005)]{Liang2005} Liang, Enwei \& Zhang, Bing. 2005, \apj, 633, 611
\bibitem[{Linsky}(1996)]{Linsky1996} Linsky, J. L. 1996, ASPC., 93, 439
\bibitem[{L$\ddot{u}$ et al.}(2012)]{Lv2012} L$\ddot{u}$ Jing, Zou Yuan-Chuan, Lei Wei-Hua. et al. 2012, \apj, 751, 49
\bibitem[{Merloni et al.}(2003)]{Merloni2003} Merloni, A., Heinz, S. \& di Matteo, T. 2003, MNRAS, 345, 1057
\bibitem[{Middleton et al.}(2013)]{Middleton2013} Middleton, M. J., Miller-Jones, James C. A., Markoff, S., Fender, R., Henze, M. et al. 2013, Nature, 493, 187
\bibitem[{Piran}(2004)]{Piran2004} Piran, T. 2004, Rev. Mod. Phys., 76, 1143
\bibitem[{Sanders et al.}(1989)]{Sanders1989} Sanders, D. B., Phinney, E. S., Neugebauer, G., Soifer, B. T., Matthews, K. 1989, \apj, 347, 29
\bibitem[{Staley et al.}(2013)]{Staley2013}   Staley, T. D., Titterington, D. J., Fender, R. P. et. al. 2013, MNRAS, 428, 3114
\bibitem[{van der Horst et al.}(2015)]{van der Horst2015} van der Horst, A. J., Levan, A. J., Pooley, G. G. et al. 2015, MNRAS, 446, 4116
\bibitem[{van den Oord}(1999)]{van1999} van den Oord, G. H. J. 1999, ASPC, 158, 189
\bibitem[{Veres et al.}(2014)]{Veres2014} Veres, Péter, Corsi, Alessandra, Frail, Dale A., Cenko, S. Bradley, \& Perley, Daniel A. 2014, arXiv1411.7368
\bibitem[Wang et al.(2014)]{Wang2014}Wang, J.-Z., Lei, W.-H., Wang, D.-X., et al., 2014, ApJ, 788, 32
\bibitem[Wu et al.(2011)]{Wu2011}Wu, Q., Zou, Y.-C., Cao, X. et al., 2011, ApJ, 740, L21
\bibitem[{Xu et al.}(2013)]{Xu2013} Xu, D., et al. 2013, GCN Circ, 14478
\bibitem[{Yonetoku et al.}(2004)]{Yonetoku2004} Yonetoku, D., Murakami, T., Nakamura, T. 2004, \apj, 609, 935
\bibitem[{Zauderer et al.}(2013)]{Zauderer2013} Zauderer, B. A., Berger, E., Margutti, R. et al. 2013, \apj, 767,161
\bibitem[{Zhang}(2007)]{Zhang2007} Zhang, B., 2007, ChJAA, 7, 1
\end{thebibliography}
\end{document}